\documentclass[12pt,preprint]{aastex}

\usepackage{epsfig}

\shorttitle{He~I Emission in H~II Regions}
\shortauthors{Porter et al.}

\begin{document}
\title{He~I Emission in the Orion Nebula and Implications for Primordial Helium Abundance}


\author{R. L. Porter, G. J. Ferland, \& K. B. MacAdam}
\affil{Dept. of Physics and Astronomy, University of Kentucky, Lexington, KY 40506}
\email{rporter@pa.uky.edu}

\begin{abstract}
We apply a recently developed theoretical model of helium emission to observations of both the Orion Nebula and a sample of extragalactic H~II regions.  In the Orion analysis, we eliminate some weak and blended lines and compare theory and observation for our reduced line list.  With our best theoretical model we find an average difference between theoretical and observed intensities $\left\langle I_{\mathrm{predicted}}/I_{\mathrm{observed}}-1\right\rangle = 6.5\%$.  We argue that both the red and blue ends of the spectrum may have been inadequately corrected for reddening.  For the $22$ highest quality lines, with $3499~\mbox{\AA}\le\lambda\le6678~\mbox{\AA}$, our best model predicts observations to an average of $3.8\%$.  We also perform an analysis of the reported observational errors and conclude they have been underestimated.
In the extragalactic analysis, we demonstrate the likelihood of a large systematic error in the reported data and discuss possible causes.  This systematic error is at least as large as the errors associated with nearly all attempts to calculate the primordial helium abundance from such observations.  Our  Orion analysis suggests that the problem does not lie in the theoretical models.  We demonstrate a correlation between equivalent width and apparent helium abundance of lines from extragalactic sources that is most likely due to underlying stellar absorption.
Finally, we present fits to collisionless case-B He~I emissivities as well as the relative contributions due to collisional excitations out of the metastable $2s\,{}^{3}\!S$ term.


\end{abstract}

\keywords{atomic data---hii regions---ISM: atoms---ISM: clouds---plasmas---Orion---extragalactic}

\section{Introduction}
In an era of precision cosmology, accurate theoretical calculations of He~I emission are essential.  In order to determine the primordial helium abundance to a relative accuracy of better than one percent, which is needed to place meaningful constraints on Big Bang Nucleosynthesis models, He~I emissivities also should be known to within a percent.  
See Bridle et al. (2003) for a discussion of the need for accurate measurement of cosmological parameters.
A number of authors have discussed the errors involved (Peimbert et al. 2003, Olive \& Skillman 2004, Izotov \& Thuan 2004).  In previous papers, we presented improved calculations, in the case-B approximation (Baker \& Menzel 1938), of He~I emissivities for a range of temperatures and densities (Porter et al. 2005, hereafter Paper 1) and for the collisionless case (Bauman et al. 2005, hereafter Paper 2).  In these papers, we predicted emissivities that differed significantly from the previous calculation of He~I emissivities (Benjamin et al. 1999, hereafter B99). Here we examine the consequences of these new emissivities upon abundance determinations.    

As a test of the model helium atom, we predict emission and compare our results with the Very Large Telescope observations (Esteban et al 2004, hereafter E04) of the Orion Nebula.  We first use the simple case-B approximation  (in a single, constant temperature, homogeneous zone) discussed in Papers 1 and 2, with parameters taken from E04.  (See Appendix~\ref{Appdx} of the present work for fits to both the recombination-only emissivities, as a function of temperature, and the collisional contributions due to excitations from the metastable $2s\,{}^{3}\!S$ term, as a function of electron density.)  As expected, there are some large differences, but most of these can be readily explained by optical depth effects not included in our simple model.  We improve the agreement significantly by selecting a small set of lines and optimizing our model by varying helium abundance, temperature, and density.  Agreement is further improved with a complete simulation, in which the case-B and constant temperature constraints are removed and the emission-line region is modeled as an extended region, with full radiative transfer effects.  In this last comparison we find agreement for $22$ highest quality He~I lines to an average of $3.8\%$.

Having validated our model helium atom via comparison with high-quality spectra of the Orion Nebula, we then turn to applying our model to primordial helium.  We investigate the observations of blue compact galaxies by Izotov \& Thuan (2004).  We find large systematic differences in the Izotov \& Thuan singly ionized helium abundances determined from different lines.  We suggest that the systematic differences are primarily due to underlying absorption, and we introduce a method by which the uncertainties involved in correcting for this effect can be minimized.  We discuss other systematic effects and conclude with a summary of our results.

The model helium atom used here is a part of the plasma simulation code CLOUDY (Ferland et al. 1998) and is described in Paper 1 and Paper 2.  For the purposes of this paper, we resolve all $nLS$ terms up to $n\leq40$ and include a series of ``collapsed'' $n$-resolved levels with $41\leq n\leq100$.  These collapsed levels allow for simpler, less CPU-intensive treatment of very highly excited levels and are discussed in detail in Paper 2.  At the finite densities considered here, this approximation introduces negligible uncertainty.    

\section{The Orion Nebula}
\subsection{Analysis of Observations}
The E04 Very Large Telescope observations of the Orion Nebula include 100 helium emission-line identifications.  We dispute the identification of $\lambda7937$ (identified by E04 as the helium line $27d\,{}^{3}\!D-3p\,{}^{3}\!P$) because of the absence in their observation of lines $nd\,{}^{3}\!D-3p\,{}^{3}\!P$ with $19\leq n\leq 26$ and suggest that the correct identification is more likely the Fe~I line at $\lambda7937.13$ (NIST Atomic Spectra Database, version 3.0.3\footnote[1]{see http://physics.nist.gov/PhysRefData/ASD/index.html}).  We note that E04 identify four other lines as Fe~I emission, although three of these are marked with a question mark to indicate an uncertain identification.

\begin{figure}
\centering
\includegraphics[keepaspectratio=true]{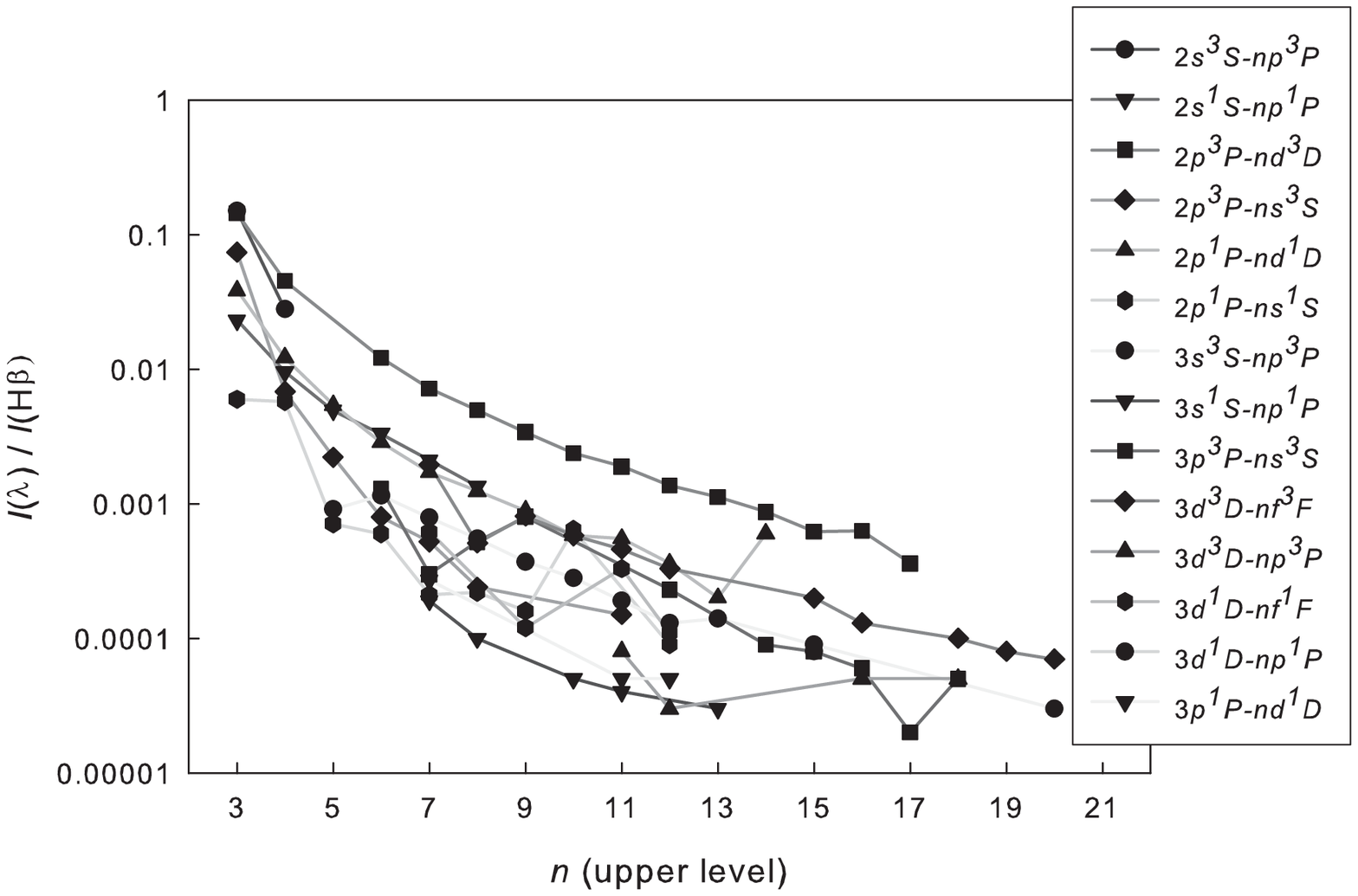}
\caption[]{Intensities of all helium lines observed by E04, relative to H$\beta$, grouped in series, and as a function of the principal quantum number of the upper level.}
\label{fig:errors}
\end{figure}

In Figure~\ref{fig:errors}, we plot the E04 reddening-corrected intensities (relative to H$\beta$) of all lines, grouped in series and as a function of the principal quantum number of the upper level.  A smooth, monotonic progression of these intensities is expected.  A cursory examination of the trends suggests that many of the weaker lines may be uncertain by a factor of 2 or more.  For the purposes of this study, we discard all lines with $I(\lambda)<0.001\times I($H$\beta)$, leaving 32 lines.  We also discard $\lambda8362$ ($6p\,{}^{3}\!P-3s\,{}^{3}\!S$) because of a possible blend with Cl~II $\lambda8361.84$.  Compared to Esteban et al. (1998) and Baldwin et al. (2000), the intensity of $\lambda5048$ ($4s\,{}^{1}\!S-2p\,{}^{1}\!P$) is anomalously high.  There is also a possible blend with Fe~II $\lambda5048.19$ and/or Fe~I $\lambda5048.43$, as evidenced by the fact that the redshift of the observed feature, if regarded as the unblended He~I line at $\lambda5048$, would be many standard deviations larger than the mean redshift of the remaining helium lines.  We therefore discard $\lambda5048$ as well.  The remaining 30 lines are given in column 1 of Table~\ref{table:casebcompare}.  (The other columns of Table~\ref{table:casebcompare} are discussed below.)  

As further evidence that some of the E04 line intensity ratios may have uncertainties larger than reported, we compare relative intensities of lines originating from the same upper level to their theoretical values.  Where optical depth effects are not important the theoretical ratios of these line intensities should be extremely accurate.  The ratio is as follows:
\begin{equation}
\frac{I(nLS-n'L'S)}{I(nLS-n''L''S)} = \frac{A(nLS-n'L'S)}{A(nLS-n''L''S)}\frac{\lambda''}{\lambda'}
\end{equation}
where $\lambda'$ and $\lambda''$ are the wavelengths of the transitions $nLS-n'L'S$ and $nLS-n''L''S$, respectively.  In Table~\ref{table:initiallevel}, we report the theoretical and observed ratios of all pairs of lines with the same upper level.  The percent difference between the theoretical and observed ratios is given, and, in the last column, we report error propagated from the errors given by E04.  
If we assume that the errors reported in E04 represent one standard deviation, we would expect (for a normal distribution) about one pair out of the 16 pairs of lines ($5\%$) to deviate by more than two standard deviations, but we find that three pairs do ($19\%$). Five or six pairs ($33\%$) would be expected to deviate by at least one standard deviation, but we find that seven pairs do ($44\%$).  Ten or eleven pairs ($68\%$) would be expected to deviate by less than one standard deviation, but we find that only nine do ($56\%$).  Little changes if we exclude the three lines (in two pairs) in Table~\ref{table:initiallevel} for which E04 do not report an error except to say that it is likely over $40\%$.  Note in particular that three of the pairs deviate by more than four standard deviations, which violates the Chebychev Inequality applicable to any statistical distribution with finite variance (Hamilton 1964).
This analysis suggests that it is likely that the errors are underestimated in E04. 
 
\begin{figure}
\centering
\includegraphics[keepaspectratio=true]{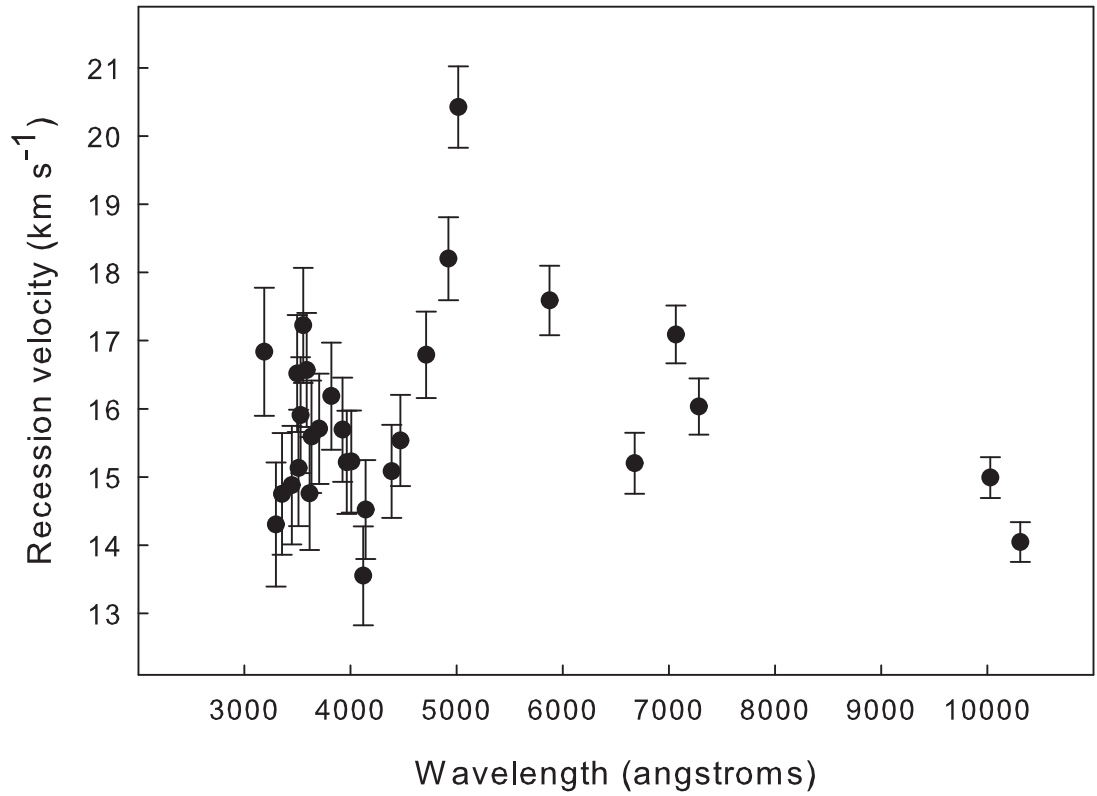}
\caption[]{Recession velocities of our reduced set of E04 helium lines plotted versus wavelength.  We have not included $\lambda3889$ because of the strong blend with H$8$.}
\label{fig:redshifts}
\end{figure}

In an attempt to confirm that our 30 lines are correctly identified and not significantly blended, we plot in Figure~\ref{fig:redshifts} heliocentric recession velocities versus wavelength.  We exclude $\lambda3889$ ($3p\,{}^{3}\!P-2s\,{}^{3}\!S$) from the plot but not from our line list because of strong blending with H$8$.  The error bars correspond to the least significant digit in the observed wavelengths, as reported by E04.  The two lines with the greatest recession velocities are $\lambda5016$ ($3p\,{}^{1}\!P-2s\,{}^{1}\!S$) and $\lambda4922$ ($4d\,{}^{1}\!D-2p\,{}^{1}\!P$), which may be inaccurate due to being near the edge of the $\lambda4750-6800$ and $\lambda3800-5000$ wavelength intervals listed in Table~1 of E04.  The average recession velocity is $15.8$~km~s$^{-1}$, and the standard deviation is $1.4$~km~s$^{-1}$.  
O'Dell (2001) reports $17.9\pm1.3$~km~s$^{-1}$ as a typical heliocentric velocity of ``medium ionization'' ions, which  include He$^{+}$.  The low end of O'Dell's reported range overlaps with the high end of the range we find here.  We conclude that there are no misidentifications or unknown blends in our line list.

\subsection{Theoretical Modeling}
\begin{figure}
\centering
\includegraphics[keepaspectratio=true]{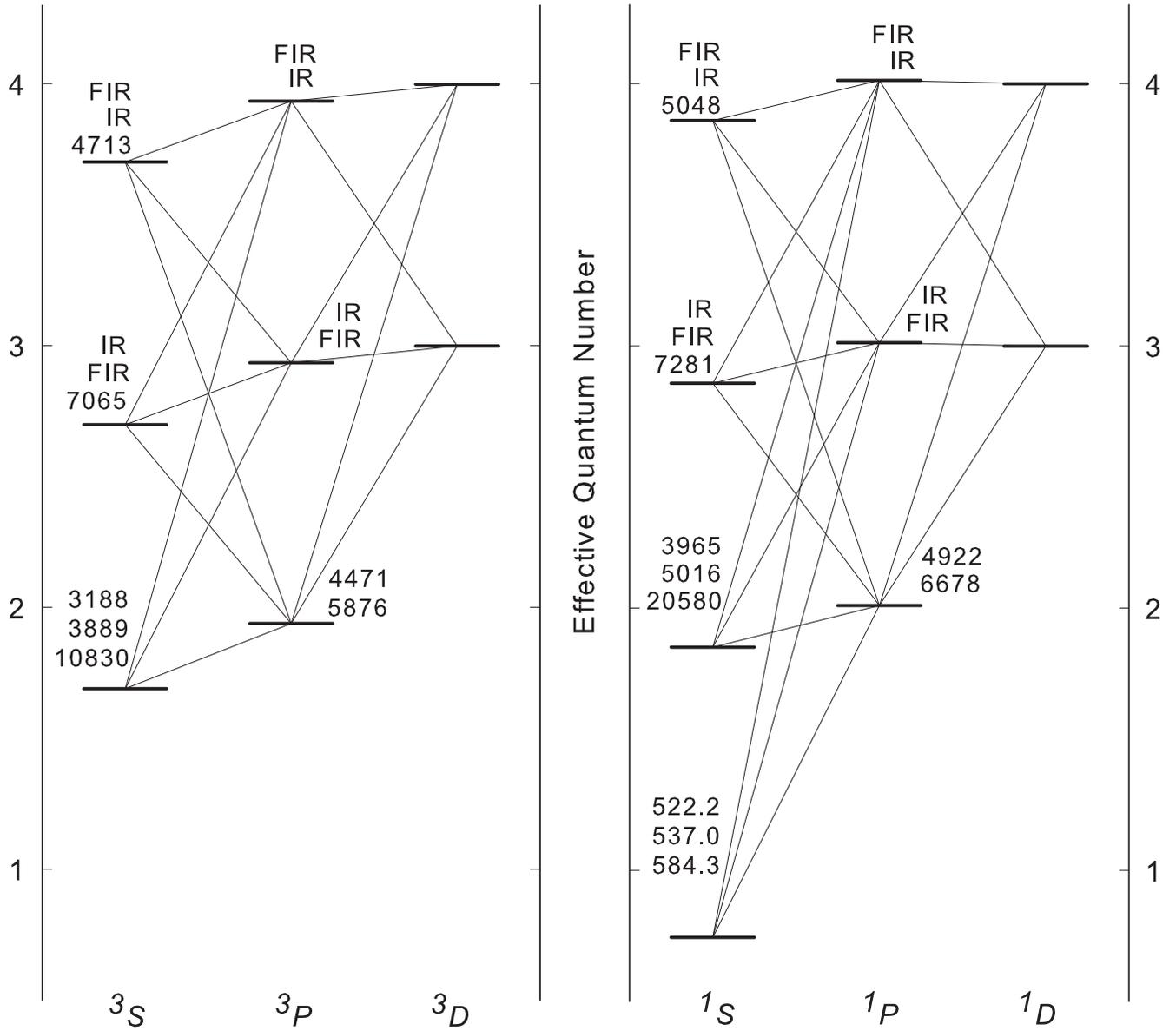}
\caption[]{Grotrian diagrams of the singlet and triplet systems of the helium atom.  The effective quantum number, $n-\delta$, is as defined in equation 1 of Paper 2.  Labels stacked near level $nl\,{}^{2S+1}\!L$ indicate the wavelengths in angstroms of transitions $nl\,{}^{2S+1}\!L-n'l'\,{}^{2S+1}\!L'$, where $l'=L'=l+1$.  The labels are arranged vertically in the same sense as the $n'l'\,{}^{2S+1}\!L'$ levels are arranged.}
\label{fig:grotrian}
\end{figure}

We calculate with CLOUDY a single, homogeneous, constant temperature, constant density, case-B model (Model I) using the helium abundance, density, and temperature derived in the E04 best-fit analysis.  In columns 2 of Table~\ref{table:casebcompare} we present the fractional difference, $I($predicted$)/I($observed$)-1$, for each line.  In column 3 we present the $\chi^2$ value for each line. Optical depth effects are important.  Among permitted triplet lines, for example, the intensities of both $\lambda3889$ and $\lambda3188$ ($4p\,{}^{3}\!P-2s\,{}^{3}\!S$) are overpredicted by case-B calculations, suggesting that, in the observed data, both lines are suffering significant self-absorption due to the metastability of $2s\,{}^{3}\!S$ (see Robbins 1968).  This argument is strengthened by the fact that the intensities of $\lambda7065$ ($3s\,{}^{3}\!S-2p\,{}^{3}\!P$) and $\lambda4713$ ($4s\,{}^{3}\!S-2p\,{}^{3}\!P$) are underpredicted in case B and apparently enhanced at the expense of the previous two lines.  (See the Grotrian diagrams of the triplet and singlet systems of helium in Figure~\ref{fig:grotrian}.)  Because the intensity of $\lambda4121$ ($5s\,{}^{3}\!S-2p\,{}^{3}\!P$) is also underpredicted in case B, it appears likely that this trend would extend to it and $\lambda2945$ ($5p\,{}^{3}\!P-2s\,{}^{3}\!S$).  However, we cannot confirm whether $\lambda2945$ is suffering self-absorption because it is outside the wavelength range of the observation.  

There is also some deviation from case-B predictions in singlet lines.  For example, the intensities of $\lambda5016$ ($3p\,{}^{1}\!P-2s\,{}^{1}\!S$) and $\lambda3965$ ($4p\,{}^{1}\!P-2s\,{}^{1}\!S$) are both overpredicted by case-B calculations.  The first and most likely explanation is that, contrary to the case-B assumption, the VUV lines to ground $\lambda537.0$ ($3p\,{}^{1}\!P-1s\,{}^{1}\!S$) and $\lambda522.2$ ($4p\,{}^{1}\!P-1s\,{}^{1}\!S$) are partially escaping the cloud rather than being completely reabsorbed.  Emission in these lines will necessarily weaken the intensities of $\lambda\lambda5016$ and $3965$ relative to case B.  A second explanation is that self-absorption effects from $2s\,{}^{1}\!S$ are important in these singlet lines in the same way they are important for their triplet counterparts $\lambda\lambda3889$ and $3188$, as discussed above.  If self-absorption were important, however, one would expect the observed intensity of $\lambda7281$ ($3s\,{}^{1}\!S-2p\,{}^{1}\!P$) to be enhanced relative to case B because each absorption of $\lambda\lambda5016$ or $3965$ will provide a new opportunity to populate $3s\,{}^{1}\!S$, necessarily enhancing $I(\lambda7281)$; we find the opposite.  This point deserves emphasis.  We calculate case-B intensities of $\lambda7281$ that are $25\%$ \textit{greater} than in the E04 observation, for which they report an uncertainty of $8\%$.  If $\lambda522.2$ partially escapes the cloud, $I(\lambda7281)$ will weaken somewhat relative to case B.  This effect, however, would also decrease emission in $\lambda6678$ ($3d\,{}^{1}\!D-2p\,{}^{1}\!P$), which we do not see.  We believe, therefore, that $I(\lambda7281)$ being overestimated by case B cannot be explained by these optical depth effects.  We address this issue again below.

Next, we modify Model I to allow CLOUDY's optimizer to vary the helium abundance, density, and temperature in an attempt to minimize the average $\chi^2$ between our predicted values and the observed values of a small set of carefully chosen lines.  We will label the new model ``Model II''.  We exclude any lines with upper principal quantum number $n_u > 5$, as well as $\lambda\lambda5016$, $7281$, and $3965$ because, as discussed above, they are strongly affected by $\lambda\lambda537.0$ and $522.2$ escaping the cloud.  In order to minimize the uncertainties involved in correcting for the optical depth effects in triplet lines, we introduce a novel approach.  We note that absorption of a photon at $\lambda3889$ will inevitably lead either to re-emission of a $\lambda3889$ photon or to decay to $3s\,{}^{3}\!S$ followed by emission of a photon at $\lambda7065$.  Either way, the total number of photons in $\lambda\lambda3889$ and $7065$ is conserved.  Furthermore, absorption of a photon at $\lambda3188$ will result in one of the following:
\begin{enumerate}
	\item re-emission of a photon at $\lambda3188$; or
	\item decay to $4s^3S$ followed by emission of a photon at $\lambda4713$; or
	\item decay to $4s^3S$ followed by emission of a photon at $\lambda3889$; or
	\item decay to $3s^3S$ followed by emission of a photon at $\lambda7065$; or
	\item decay to $3d^3D$ followed by emission of a photon at $\lambda\lambda5876$, $3889$, or $7065$.  
\end{enumerate}
Therefore, the sum of photons in $\lambda\lambda3188$, $3889$, $4713$, $5876$, and $7065$ is independent of the optical depth $\tau_{3889}$ (and $\tau_{3188}$ as well, since $\tau_{3889}$ and $\tau_{3188}$ are related by atomic data alone).  We include the hydrogen line H$8$ at $\lambda3889$ to minimize the uncertainty of deblending this line from the close helium line.  We use this photon sum, referred to as ``triplet photon sum'' in Table~\ref{table:casebcompare}, in our optimization and exclude from our optimization each of the individual lines included in the photon sum. (Note that in the vast majority of spectra of H~II regions, $\lambda3188$ is not observed).  

\begin{figure}
\centering
\includegraphics[keepaspectratio=true]{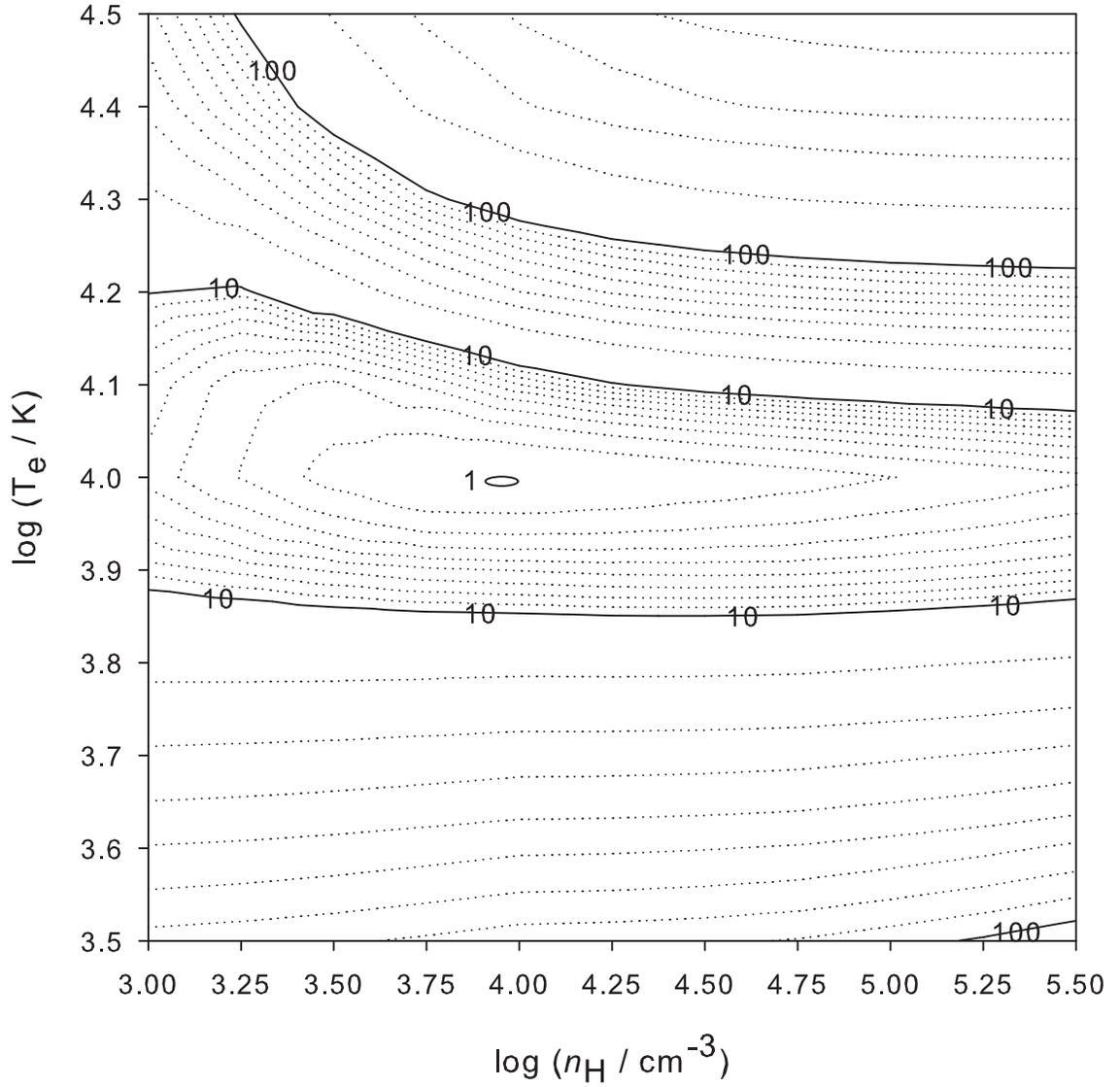}
\caption[]{Average $\chi^2$ between seven predicted and observed quantities as a function of hydrogen density, $n_{\mathrm{H}}$, and electron temperature, $T_e$.}
\label{fig:chi2}
\end{figure}

We are left with the following seven quantities to be optimized:  the intensities relative to H$\beta$ of $\lambda\lambda6678$, $4922$, $3614$ ($5p\,{}^{1}\!P-2s\,{}^{1}\!S$), $4388$ ($5d\,{}^{1}\!D-2p\,{}^{1}\!P$), $4471$ ($4d\,{}^{3}\!D-2p\,{}^{3}\!P$), and $4121$; and the triplet photon sum discussed above.  We weight each value by an uncertainty.  For the line intensities, we use the uncertainties given by E04.  For the photon sum, we calculate the average of the uncertainties of the individual lines, weighted by photon count, and obtain $4\%$.  
The optimizer finds best-fit values for the helium abundance ($y^{+}=$~He$^+$/H$^+$, by number), log electron density (cm$^{-3}$), and temperature equal to $0.0874\pm0.0005$, $3.94^{+1.1}_{-0.4}$, and $9800\pm900$~K, respectively. The values found in E04 are $0.0874\pm0.0006$, $3.95\pm0.01$, and $8730\pm320$~K, respectively.  Note that our values were determined using only helium line intensities relative to H$\beta$ while E04 determined temperature and density from a number of diagnostics that did not include helium intensities at all.  Figure~\ref{fig:chi2} is a contour plot of the average $\chi^2$ between the seven predicted and observed quantities as a function of density and temperature, with the helium abundance fixed at the above optimal value.  The $\chi^2$ values in this plot have been scaled so that the minimum is at unity.  The temperature is more tightly constrained than the density, which has a rather large standard deviation. 

We substitute our new optimal values into Model I to evaluate Model II, the results of which are presented in columns 4 and 5 of Table~\ref{table:casebcompare}.   The last row of Table~\ref{table:casebcompare} contains the average $\chi^2$ for each model.  The average $\chi^2$ obtained with Model II is roughly half of the average obtained with Model I.  
  
Finally, we again use CLOUDY, this time calculating a more realistic model (Model III).  This model is a plane-parallel slab with abundances typical of H~II regions (Baldwin et al. 1991; Rubin et al. 1991; Osterbrock et al. 1992) and grain distributions characteristic of those found in the Orion Nebula (Baldwin et al. 1991; van Hoof et al. 2004).  The slab is heated by a star with a characteristic temperature of 39,600 K (Kurucz 1979).  We set the surface flux of hydrogen-ionizing photons, $\Phi($H$) = 10^{13}$~cm$^{-2}$~s$^{-1}$ and include a microturbulence velocity field parameter equal to $16$~km~s$^{-1}$, the average recession velocity found above.  We enforce constant pressure and include the cosmic ray background.  The case-B constraint is not specified, and radiative transfer effects (including continuum fluorescence, line destruction by background opacities, and line optical depth) are treated self-consistently.  The results of Model III are presented in columns 6 and 7 of Table~\ref{table:casebcompare} and discussed below.  

\subsection{Discussion}
\begin{figure}
\centering
\includegraphics[keepaspectratio=true]{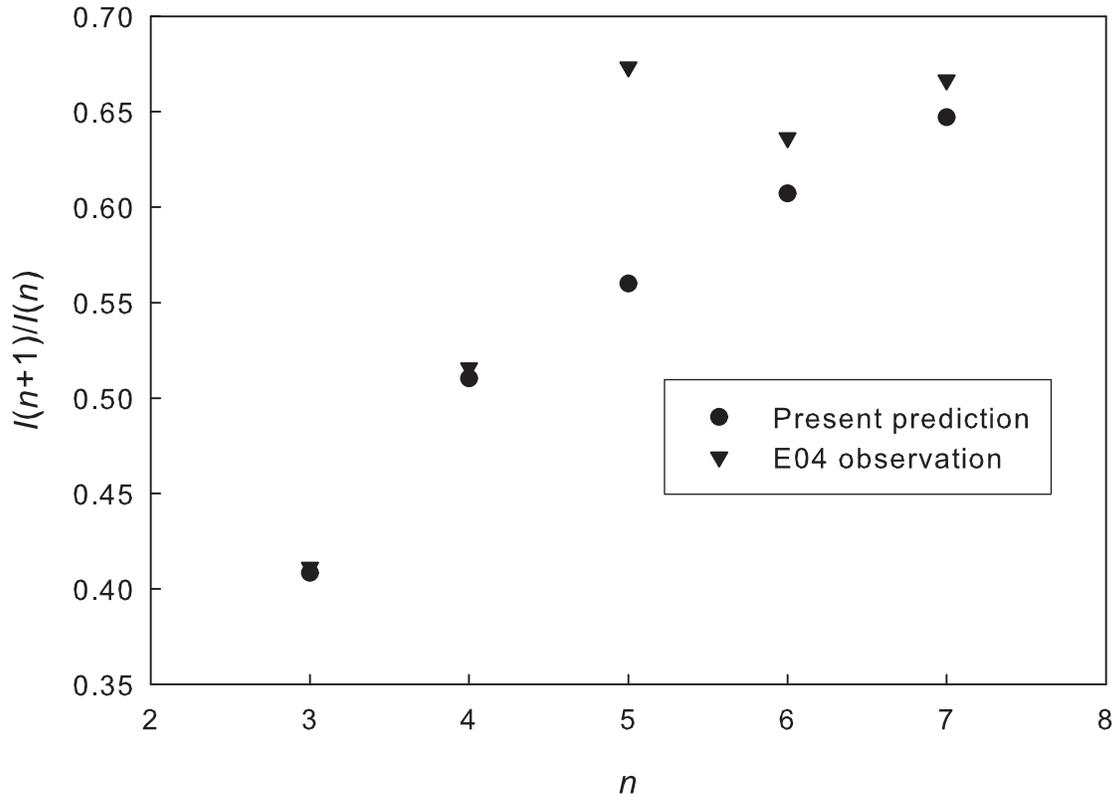}
\caption[]{Ratio $I(n+1)/I(n)$ of intensities of lines in the $np\,{}^{1}\!P-2s\,{}^{1}\!S$ series, plotted versus the principal quantum number $n$ of the upper level, for both the E04 observations and the present (Model III) predictions.  The observed ratio appears anomalously high at $n=5$.}
\label{fig:deriv}
\end{figure}

The average $\chi^2$ achieved with our Model III is dramatically better than with either Model I or Model II.  However, several problems remain.  The four lines in our set with the shortest wavelengths are systematically underpredicted by $16-17\%$.  One of them, $\lambda3188$, could possibly be explained by optical depth effects, although this seems unlikely since $I(\lambda3889)$ is accurately predicted.  The other three lines are $\lambda\lambda3448$, $3355$, and $3297$, which are in the same series: $np\,{}^{1}\!P-2s\,{}^{1}\!S$ (for $n=6$, $7$, and $8$).  For these three lines and the rest of the observed $np\,{}^{1}\!P-2s\,{}^{1}\!S$ series, we plot in Figure~\ref{fig:deriv} the ratio $I(n+1)/I(n)$ of both the predicted and observed intensities as a function of the principal quantum number $n$ of the upper level.  The ratios of the predicted values are smoothly increasing, as expected, but the ratios of the observed values are anomalously high at $n=5$, suggesting that the $\approx16\%$ difference noted above is due to inaccuracies in the E04 results.  

Blagrave et al. (2006) have performed a thorough study of reddening toward the Orion Nebula, using much higher quality, recent observations.  They suggest that the linear extrapolation of the Costero \& Peimbert (1970) reddening function that E04 employ may significantly overestimate the extinction shortward of $\lambda3500$.  We support this conclusion.  The new extinction corrections calculated by Blagrave et al. (based on the analytical extinction laws of Cardelli et al. 1989) would also work to reduce or eliminate the discrepancies between the present results and E04 for wavelengths longward of about $\lambda7000$ (see Figure~7 of Blagrave et al. 2006).  We suggest, in particular, that the poor agreement we have shown for $\lambda7281$ may be primarily due to inaccurate reddening corrections.  The average difference $I_{\mathrm{predicted}}/I_{\mathrm{observed}}-1$ of the lines in Table~\ref{table:casebcompare} is $6.5\%$.  If we consider only the $22$ lines with $3499~\mbox{\AA}\le\lambda\le6678~\mbox{\AA}$, we find an average difference of $3.8\%$.

Our comparison between theoretical and observed helium intensities is limited primarily by the uncertainty in the observations.  After eliminating some blends and lines with poor SNR, we have demonstrated good agreement between theoretical and observed values of over $20$ helium line intensities in the Orion Nebula.  
We believe our theoretical model is capable of accurately predicting many more line intensities, including lines originating from larger principal quantum number.  

\section{Primordial Helium Abundance}

\subsection{Analysis of Observations}
Recently, Izotov \& Thuan (2004, hereafter IT04) presented extensive data from 33 observations of blue compact galaxies (BCGs), and used a subset of the strongest He~I lines to estimate the helium abundance of each system.  While IT04 also included in their final analysis observations from Izotov \& Thuan (1998), Izotov et al. (2001), Guseva et al. (2003a), and Guseva et al. (2003b), we consider (for simplicity) only the 33 new observations of IT04 in the present analysis.  

\begin{figure}
\centering
\includegraphics[scale=.8]{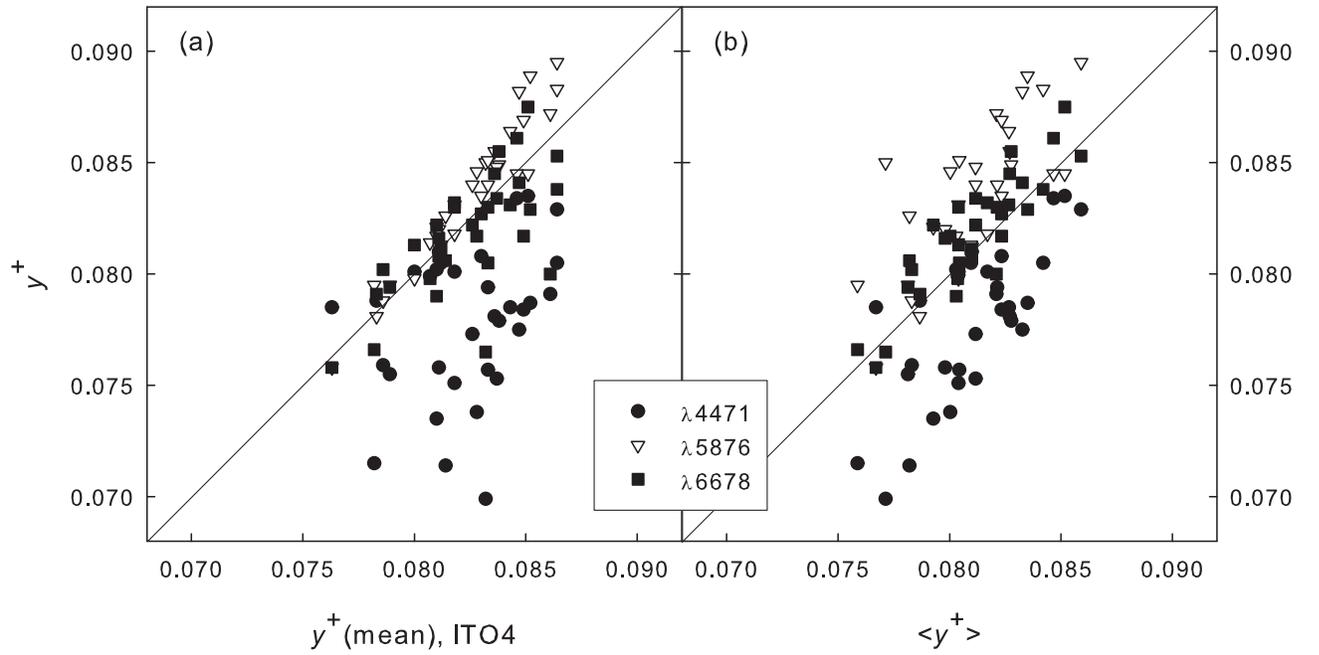}
\caption[]{IT04 values of $y^{+}$ determined from $\lambda\lambda4471$, $5876$, and $6678$ versus (a) weighted mean $y^{+}(mean)$ from Table~4 of IT04, and (b) a simple mean, denoted by $<y^{+}>$.}
\label{fig:yplus}
\end{figure}

We examined the values of $y^{+}(\lambda4471)$, $y^{+}(\lambda5876)$, and $y^{+}(\lambda6678)$ reported by IT04 in their Table~4 and found statistically significant systematic differences.  The average and standard deviation of each of these quantities are presented in Table~\ref{table:IT04abundances}.  The left panel of Figure~\ref{fig:yplus} plots these values versus the weighted means, $y^{+}($mean$)$, defined in equation 2 of IT04.  The right panel plots the same values versus the simple means.  The left panel demonstrates how skewed the weighted means are toward $y^{+}(\lambda5876)$ (because of the much stronger signal of $\lambda5876$) but gives the superficial impression that there is much less deviation in the $y^{+}(\lambda5876)$ values than in the $y^{+}(\lambda6678)$ and $y^{+}(\lambda4471)$ values.  The right panel preserves the relative deviations in each set.  The weighted and simple means differ by as much as $8\%$.
This analysis is similar to an analysis performed by Skillman et al. (1998) on the Izotov et al. (1997) dataset.

\subsection{Sources of Uncertainty}
IT04 derived their abundances using the emissivities of B99, and we find trends similar to those shown in Figure~\ref{fig:yplus} when using case-B emissivities from CLOUDY.  The largest differences are between the abundances determined from $\lambda4471$ and those determined from $\lambda\lambda5876$ and $6678$, amounting to about $7$ and $5\%$, respectively.  Note that $\lambda\lambda5876$ and $6678$ are lines originating from \textit{yrast}\footnote[2]{In regular use in physics literature but rarely in astronomy, the terms \textit{yrare} and \textit{yrast} (introduced by Grover 1967) are comparative and superlative modifications of the Swedish word \textit{yr}, meaning ``dizzy", and refer, respectively, to levels having high angular momentum and the highest angular momentum (for a given principal quantum number).} levels.  Theoretical emissivities from yrast levels may be systematically underestimated in the low-density limit by $\approx1-2\%$ due to the inaccuracy of modelling the helium atom with a finite number of levels (see Paper II).  This effect is negligible at finite densities as low as $100$~cm$^{-3}$ as collisions force highly excited states to local thermodynamic equilibrium and is also too small to explain the systematic differences found in $y^{+}$ values.  Other effects must be involved.

One possible explanation is the use of inaccurate reddening corrections.  IT04 use the interstellar reddening function of Whitford (1958).  Olive \& Skillman (2004) compared the extinction laws of Cardelli et al. (1989) and Whitford in their Figure~2, which shows that use of the Cardelli et al. extinction law instead of the Whitford extinction law would increase the (reddening-corrected) intensity $I(\lambda6678)$ by about $1.5\%$ relative to the $I(\lambda5876)$, thereby increasing the helium abundance necessary to produce the $\lambda6678$ emission (again, relative to $\lambda5876$).  This change would nearly eliminate the systematic difference between the helium abundances determined from $\lambda\lambda6678$ and $5876$, but the systematic difference between the abundances determined from these two lines and $\lambda4471$ would remain.  The uncertainties involved in the reddening correction can be minimized by deriving $y^{+}$ values from a pair of helium and hydrogen lines with a small wavelength difference.  We calculated $y^{+}(\lambda6678)$ by referencing $\lambda6678$ to H$\alpha$ ($\lambda6563$) instead of H$\beta$ ($\lambda4861$).  Similarly, we calculated $y^{+}(\lambda4471)$ and $y^{+}(\lambda5876)$ by referencing the lines to H$\gamma$ ($\lambda4340$) and H$\alpha$ ($\lambda6563$), respectively.  The $y^{+}$ values obtained are less than the corresponding values derived when referencing the lines to H$\beta$, with $y^{+}(\lambda5876)$ and $y^{+}(\lambda6678)$ each decreased by $0.6\%$ and $y^{+}(\lambda4471)$ decreased by $1.4\%$.  These results are not consistent with the changes produced by using the Cardelli et al. law instead of the Whitford law, but both changing the reference line and changing the extinction law suggest uncertainties of about $1-2\%$.  While certainly significant in primordial helium calculations, we find no evidence that reddening corrections are responsible for the large systematic differences discussed above.   

For a small subset of systems, IT04 dramatically reduce the systematic differences in $y^{+}$ values by accounting for underlying stellar absorption, using theoretical absorption equivalent widths ($EW_a$) from Gonz\'alez Delgado et al. (1999) and making reasonable estimates for any $EW_a$ not calculated in that work.  Their method does not allow for variations in stellar population (i.e., they apply the same set of $EW_a$ to every galaxy).    
The need to allow for $EW_a$ to vary for different targets was demonstrated by Olive \& Skillman (2001) and applied to the IT04 data in Olive \& Skillman (2004).
IT04 find that after changing $EW_a(\lambda4471)$ from $0.4~\mbox{\AA}$ to $0.5~\mbox{\AA}$ their minimization procedure results in an $1\%$ increase in the primordial helium abundance $Y_p$, yet they report uncertainies of less than $1\%$ on each of their $Y_p$ values.  Allowing variations between galaxies would surely increase this uncertainty further.  

Recently, Fukugita \& Kawasaki (2006) reanalysed the entire new set of IT04 observations by considering underlying stellar absorption of the helium lines.  It is important to note that while Fukugita \& Kawasaki do significantly reduce $\chi^2$ in the relation between helium abundance and metallicity ($dY/dZ$), they do not actually model the stellar absorption.  Instead they introduce a free parameter that serves to modify the equivalent width of a line.  Contrary to the analysis performed by IT04, Fukugita \& Kawasaki use the same absorption equivalent width for each helium line, although (again contrary to IT04) they use a different absorption equivalent width for each galaxy.  Comparison with a theoretical model such as that of Gonz\'alez Delgado et al. would be beneficial for constraining the Fukugita \& Kawasaki analysis.  See Tremonti et al. (2004) for an application of this technique.

\begin{figure}
\centering
\includegraphics[keepaspectratio=true]{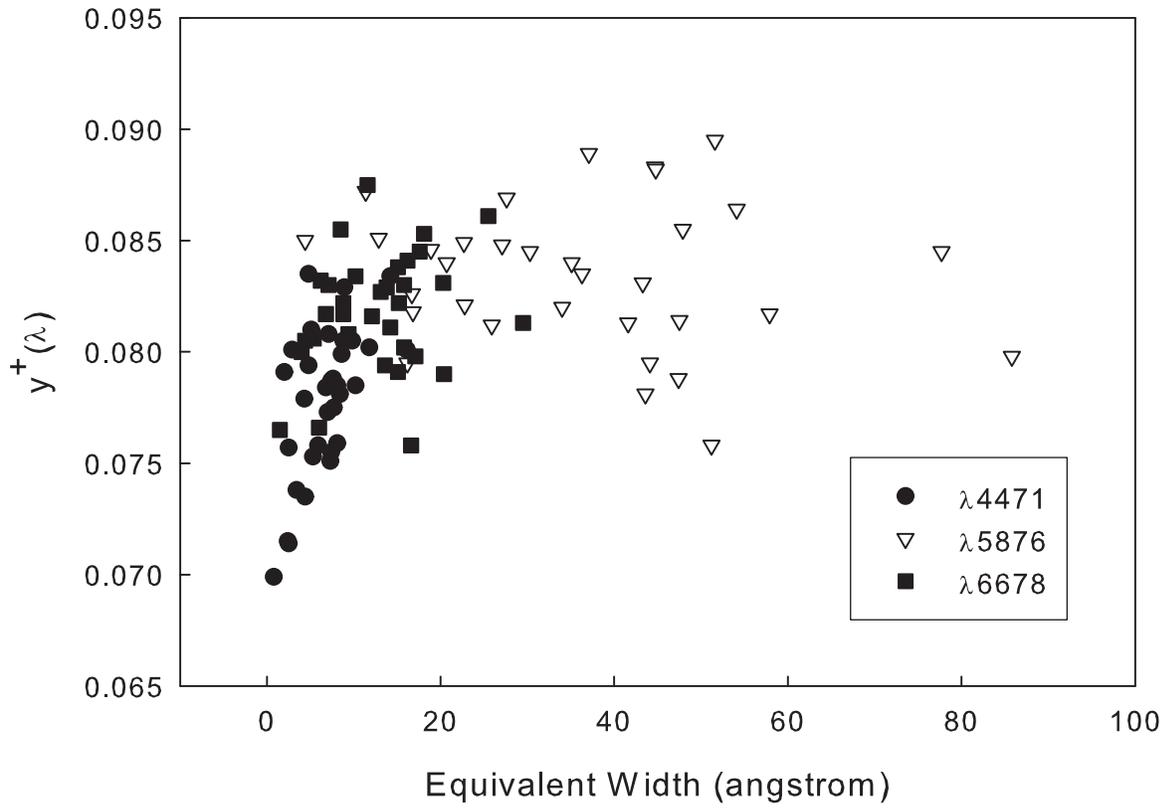}
\caption[]{IT04 values of $y^{+}$ determined from $\lambda\lambda4471$, $5876$, and $6678$ versus equivalent width.}
\label{fig:equivwidth}
\end{figure}

Figure~\ref{fig:equivwidth} plots the $y^{+}(\lambda)$ values that IT04 determined from $\lambda\lambda4471$, $5876$ and $6678$ (without correcting for underlying stellar absorption of He~I lines) versus the equivalent width of the line.  Lower $y^{+}(\lambda)$ values tend to correlate with low equivalent width.  (A similar analysis with a different dataset has been performed in Skillman et al. 1998 with the same result.)  The trend suggests that excluding lines with an equivalent width less than, say, $10~\mbox{\AA}$ in primordial helium analyses could reduce the uncertainties involved in correcting for underlying stellar absorption.  Both Olive \& Skillman (2004) and IT04 advocate selecting targets using an equivalent width cutoff but only with respect to the equivalent width of H$\beta$.  By applying a similar \textit{additional} cutoff to the helium lines, one could, in theory, reduce the uncertainty due to underlying stellar absorption to an arbitrarily small amount.  We note that although a similar effect is accomplished by giving less weight to weaker lines, as is done by IT04, that method has the disadvantage that it requires sometimes large corrections with unknown uncertainties.  The simple mean of all $99$ $y^{+}(\lambda)$ values in Table~4 of IT04 is $0.0810\pm0.0039$.  If we exclude lines with equivalent width less than $10~\mbox{\AA}$, the mean is $0.0826\pm0.0034$, about $2\%$ larger than in the full set.  This procedure has allowed us to trade large \textit{systematic} uncertainties for smaller \textit{random} uncertainties, the latter of which can always be reduced again by making more observations.  It is important to note, however, that, irrespective of underlying stellar absorption considerations, large systematic uncertainties still dominate (see, for example, Peimbert et al. 2002).

\subsection{Discussion}
We believe that the uncertainties involved in calculating the helium abundance in extragalactic systems are much larger than are generally stated.  IT04 assign less than $1\%$ uncertainty to their value of the primordial helium abundance, but the average $y^{+}$ derived from $\lambda5876$ alone is $7\%$ greater than the average derived from $\lambda4471$ alone.  This discrepancy is large, statistically significant, and obtained using both B99 emissivities and emissivities predicted by CLOUDY.  In our analysis of the E04 Orion Nebula observations, we obtain an electron density of nearly  $9000$~cm$^{-3}$.  The BCGs observed by IT04 are less dense by one to three orders of magnitude.  The uncertainties in the theoretical emissivities in such rarified environments are dramatically less than in those found in the Orion Nebula.  We therefore believe that the present Orion analysis suggests that the systematic discrepancy in $y^{+}(\lambda)$ values discussed above is not due to problems with theoretical emissivities.  We believe that underlying stellar absorption is the largest source of uncertainty.

\section{Conclusions}

\begin{description}
\item[] We have demonstrated agreement between theory and observation (of the Orion Nebula) to an average difference of $3.8\%$ for the $22$ lines we designate as having the highest quality.

\item[] Higher quality observations of the Orion Nebula could improve agreement between theory and observations.

\item[] We support the Blagrave et al. (2006) conclusion that extrapolations of the Costero \& Peimbert (1970) reddening function shortward of $\lambda3500$ overestimate extinction toward the Orion Nebula.  We believe the corrections are $\approx15\%$ too large.  We also believe, based upon the Blagrave et al. work, that Costero \& Peimbert may have underestimated extinction longward of about $\lambda7000$.

\item[] There is a systematic uncertainty in the IT04 $y^{+}$ values derived from extragalactic sources that is most likely due to underlying stellar absorption.  The uncertainties involved in the correction may be minimized by selecting lines with large equivalent width.

\item[] Much higher quality spectra are essential in further attempts to calculate the primordial helium abundance to the accuracy required for tests of Big Bang Nucleosynthesis.
\end{description}

We thank the National Science Foundation for support through award AST 06-07028 and NASA through NNG05GG04G.  We also thank referee E. D. Skillman for his careful review of our original manuscript.

\clearpage

\appendix
\section{Fits to He~I Emissivities}
\label{Appdx}

While calculating emissivities via the above equations sidesteps the self-consistency of the CLOUDY calculations, analytical expressions for emissivities are useful in efforts to find optimal plasma parameters corresponding to particular observations.  
In Table~\ref{table:emissfits}, we present four-parameter fits to the case-B, collisionless emissivity of 33 of the strongest He~I lines.  The emissivity, $4\pi j_{\lambda}/n_{e}n_{\mathrm{He}^+}$, is calculated as follows:
\begin{equation}
\frac{4\pi j_{\lambda}}{n_{e}n_{\mathrm{He}^+}} = [a + b(\ln T_e)^2 + c\ln T_e + d/\ln T_e]\times{T_e}^{-1} \times 10^{-25}\ \ [\mathrm{ergs~cm}^{3}~\mathrm{s}^{-1}].
\label{eqn:recombpart}
\end{equation}
The fits in Table~\ref{table:emissfits} introduce negligible new errors, accurately reproducing CLOUDY predictions to better than $0.03\%$ for $5000\le T($K$)\le25000$.

Collisional contributions to these emissivities are calculated in the same manner as Kingdon \& Ferland (1995).  In Table~\ref{table:collcontrib}, we present the parameters for fits to the collisional contributions to the emissivities of any line with the given upper level.  We note that collisional excitations directly affect \textit{populations} of levels, indirectly affect emission, and are independent of the lower level of the enhanced line.  The collisional contribution, $C/R$, is calculated as follows:

\begin{equation}
\frac{C}{R} = (1 + 3552~t_4^{-0.55}/n_{e})^{-1}\times \sum\limits_{i}{a_i~t_4^{b_i}~\exp(c_i/t_4)},
\label{eqn:collispart}
\end{equation}
where $t_4$ is $T_e/10000$, and $i$ is an index that varies from $1$ to the number of terms used in the fit.  As in Kingdon \& Ferland (1995), terms comprising less than 1\% of the total are ignored here.  To find the total emissivity of a given line, simply multiply the result obtained in Equation~\ref{eqn:recombpart} by the quantity $1+C/R$ obtained in Equation~\ref{eqn:collispart} via parameters listed in Table~\ref{table:collcontrib}.  

\begin{figure}
\centering
\includegraphics[keepaspectratio=true]{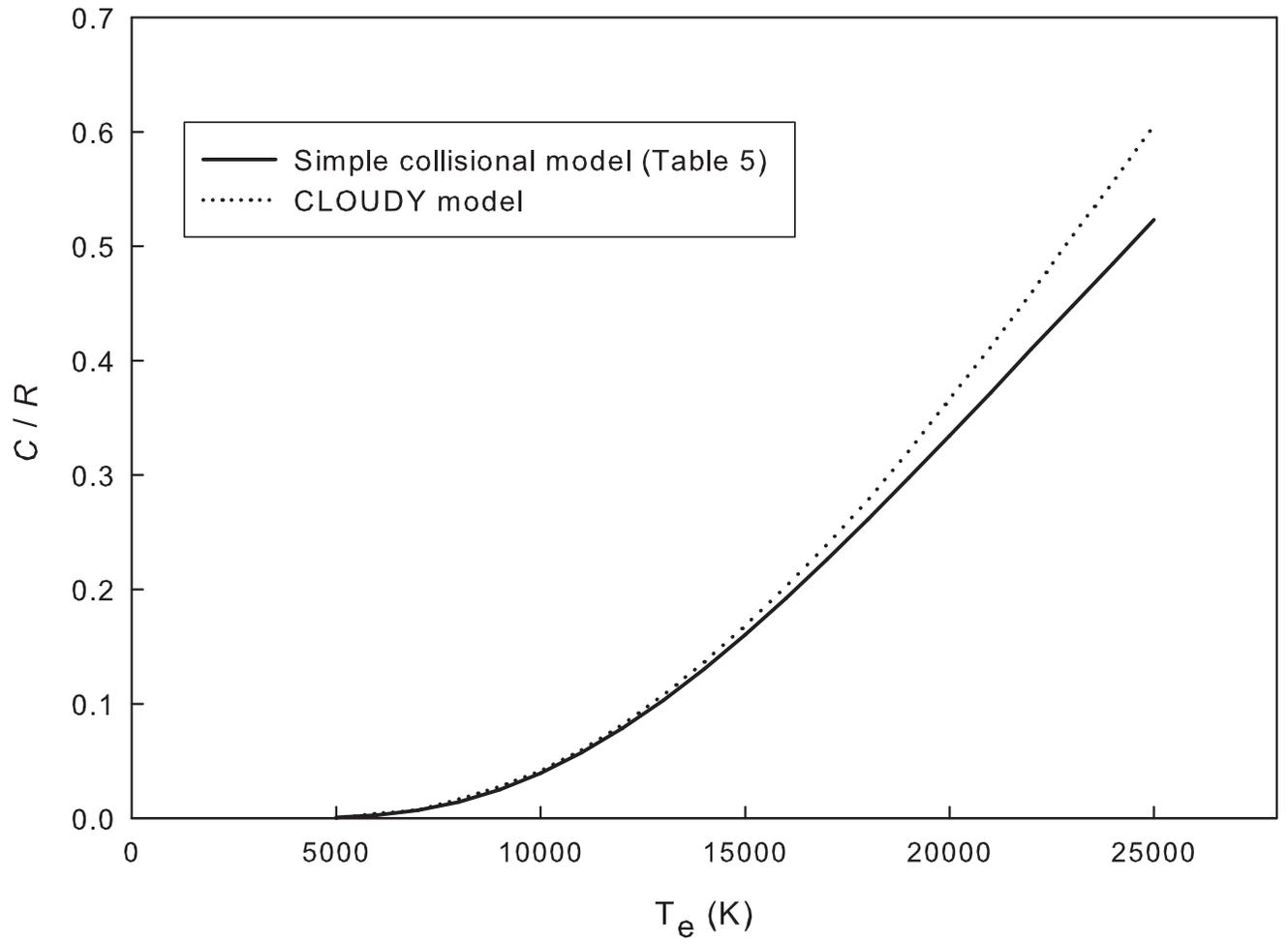}
\caption[]{Ratio of collisional to recombination contributions, $C/R$, to the emissivity of $\lambda$5876 at $n_e = 1000$~cm$^{-3}$ as a function of temperature.}
\label{fig:constant_ne}
\end{figure}

\begin{figure}
\centering
\includegraphics[keepaspectratio=true]{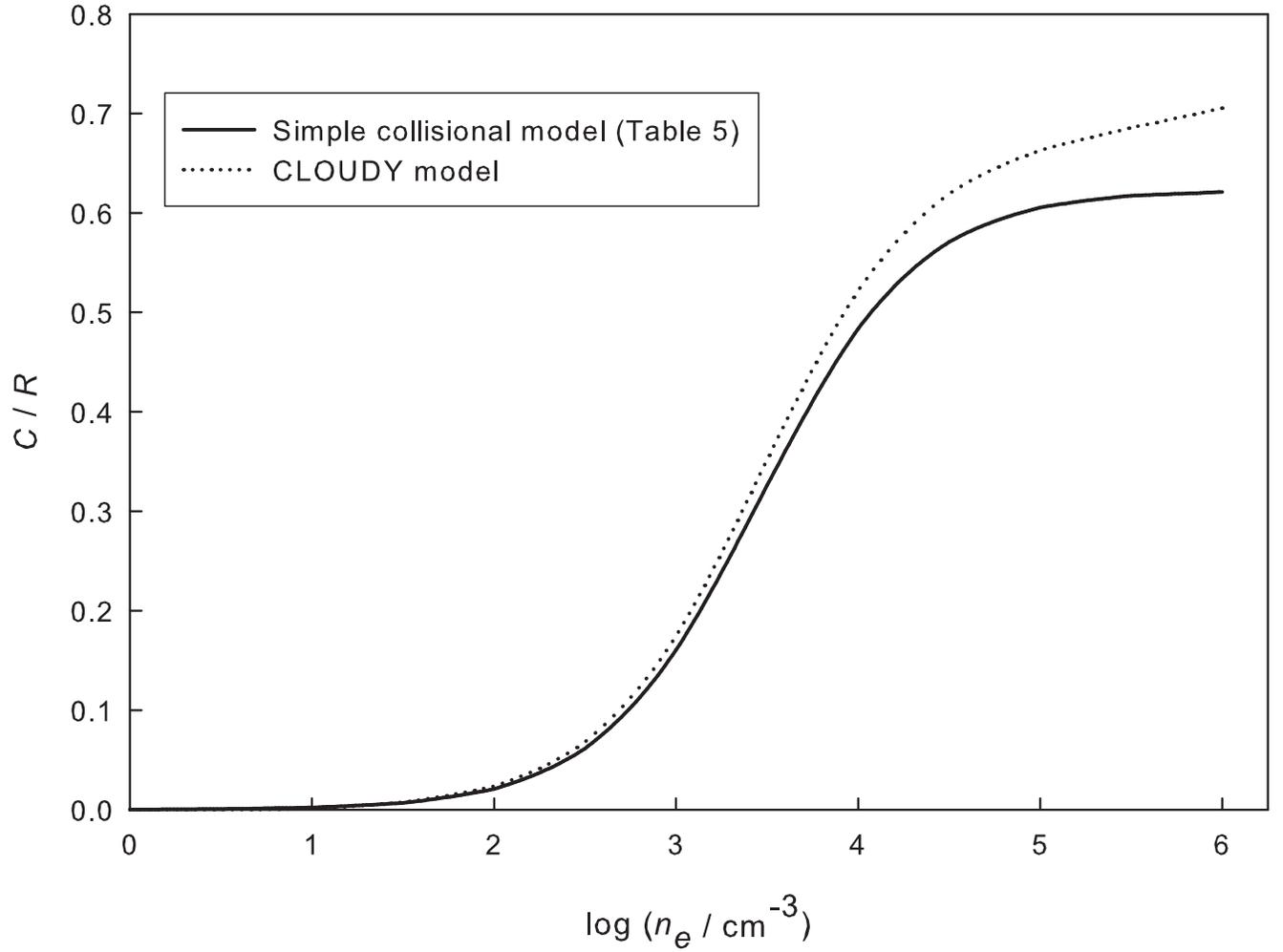}
\caption[]{Ratio of collisional to recombination contributions, $C/R$, to the emissivity of $\lambda$5876 at $T_e = 15000$~K as a function of electron density.}
\label{fig:constant_Te}
\end{figure}

Figures \ref{fig:constant_ne} and \ref{fig:constant_Te} compare CLOUDY predictions of collisional contributions to the emissivity of $\lambda$5876 with the simple fits of contributions due to excitations from $2s\,{}^{3}\!S$, as a function of temperature at a fixed density, and as a function of density at a fixed temperature.  The figures indicate that the quantities $1+C/R$ introduce negligible uncertainties at low temperatures and densities, but underestimate the collisional contribution by as much as $5\%$ at high temperatures and densities due to collisional excitations from other levels.  At such extreme conditions the collisional contributions are themselves uncertain by at least $5\%$.  At conditions typical of H~II regions ($n_e \approx 100$~cm$^{-3}$ and $T_e\approx15000$~K), the error in the collisional contribution given by Equation~\ref{eqn:collispart} amounts to a fraction of a percent of the total emissivity of $\lambda5876$.

\clearpage

\clearpage

\clearpage

\clearpage

\clearpage

\clearpage

\clearpage

\clearpage

\clearpage

\clearpage
\begin{deluxetable}{rrrrrrrrrr}
\tabletypesize{\scriptsize}
\tablecaption{Predicted intensities, relative to H$\beta$, and fractional differences, $I($predicted$)/I($observed$)-1$, and $\chi^2$ values between the line intensities calculated in each of three CLOUDY models and observed by E04.  The last row gives the average $\chi^2$ obtained for each model.}
\tablewidth{0pt}
\tablehead
{
  \colhead{Wavelength (Air)} & 
  \multicolumn{3}{c}{Model I} & 
  \multicolumn{3}{c}{Model II} & 
  \multicolumn{3}{c}{Model III} \\
  \colhead{$\mbox{\AA}$} & 
  \colhead{$I(\lambda)/I(H\beta)$} & 
  \colhead{Diff} & 
  \colhead{$\chi^2$} &
  \colhead{$I(\lambda)/I(H\beta)$} & 
  \colhead{Diff} & 
  \colhead{$\chi^2$} &
  \colhead{$I(\lambda)/I(H\beta)$} & 
  \colhead{Diff} & 
  \colhead{$\chi^2$}
}
\startdata
3188\tablenotemark{a}	&	0.041	&	0.456	&	32.542	&	0.042	&	0.520	&	42.197	&	0.023	&	-0.175	&	7.055	\\
3297\tablenotemark{a}	&	0.001	&	-0.159	&	0.394	&	0.001	&	-0.146	&	0.324	&	0.001	&	-0.168	&	0.454	\\
3355\tablenotemark{a}	&	0.002	&	-0.187	&	3.136	&	0.002	&	-0.176	&	2.707	&	0.002	&	-0.170	&	2.466	\\
3448\tablenotemark{a}	&	0.003	&	-0.177	&	5.695	&	0.003	&	-0.164	&	4.783	&	0.003	&	-0.156	&	4.239	\\
3499	&	0.001	&	-0.051	&	0.072	&	0.001	&	-0.058	&	0.095	&	0.001	&	-0.051	&	0.072	\\
3513	&	0.001	&	-0.015	&	0.008	&	0.001	&	-0.022	&	0.017	&	0.001	&	-0.015	&	0.008	\\
3530	&	0.002	&	-0.072	&	0.189	&	0.002	&	-0.078	&	0.223	&	0.002	&	-0.071	&	0.180	\\
3554	&	0.002	&	0.006	&	0.003	&	0.002	&	0.001	&	0.000	&	0.002	&	0.010	&	0.008	\\
3587	&	0.003	&	-0.032	&	0.135	&	0.003	&	-0.037	&	0.180	&	0.003	&	-0.027	&	0.093	\\
3614	&	0.005	&	-0.002	&	0.001	&	0.005	&	0.017	&	0.061	&	0.005	&	0.009	&	0.018	\\
3634	&	0.005	&	-0.045	&	0.458	&	0.005	&	-0.050	&	0.563	&	0.005	&	-0.038	&	0.315	\\
3705	&	0.007	&	-0.003	&	0.003	&	0.007	&	-0.008	&	0.027	&	0.007	&	0.008	&	0.028	\\
3820	&	0.012	&	-0.042	&	2.140	&	0.012	&	-0.048	&	2.802	&	0.012	&	-0.026	&	0.816	\\
3889\tablenotemark{b}	&	0.217	&	0.446	&	221.455	&	0.224	&	0.490	&	266.437	&	0.154	&	0.021	&	0.513	\\
3927	&	0.001	&	-0.006	&	0.009	&	0.001	&	-0.011	&	0.027	&	0.001	&	-0.007	&	0.011	\\
3965	&	0.010	&	0.055	&	3.301	&	0.010	&	0.077	&	6.506	&	0.010	&	0.023	&	0.591	\\
4009	&	0.002	&	0.095	&	3.590	&	0.002	&	0.088	&	3.119	&	0.002	&	0.091	&	3.329	\\
4121	&	0.002	&	-0.158	&	22.130	&	0.002	&	-0.070	&	3.507	&	0.002	&	0.007	&	0.033	\\
4144	&	0.003	&	0.074	&	3.458	&	0.003	&	0.067	&	2.837	&	0.003	&	0.070	&	3.047	\\
4388	&	0.006	&	0.037	&	3.370	&	0.006	&	0.032	&	2.488	&	0.006	&	0.032	&	2.547	\\
4471	&	0.045	&	-0.002	&	0.040	&	0.045	&	-0.002	&	0.040	&	0.047	&	0.033	&	10.563	\\
4713	&	0.005	&	-0.258	&	1206.619	&	0.006	&	-0.171	&	427.810	&	0.007	&	0.070	&	48.490	\\
4922	&	0.012	&	-0.007	&	0.434	&	0.012	&	-0.014	&	1.990	&	0.012	&	-0.013	&	1.760	\\
5016	&	0.026	&	0.121	&	146.383	&	0.027	&	0.150	&	223.831	&	0.024	&	0.040	&	15.917	\\
5876	&	0.128	&	-0.115	&	18.805	&	0.129	&	-0.103	&	14.755	&	0.134	&	-0.071	&	6.575	\\
6678	&	0.035	&	-0.093	&	2.887	&	0.035	&	-0.103	&	3.697	&	0.035	&	-0.100	&	3.433	\\
7065\tablenotemark{c}	&	0.034	&	-0.544	&	289.341	&	0.039	&	-0.467	&	157.203	&	0.082	&	0.115	&	2.694	\\
7281\tablenotemark{c}	&	0.007	&	0.252	&	9.904	&	0.008	&	0.375	&	21.997	&	0.008	&	0.287	&	12.879	\\
10027\tablenotemark{c}	&	0.002	&	0.089	&	0.311	&	0.002	&	0.059	&	0.135	&	0.002	&	0.073	&	0.209	\\
10311\tablenotemark{c}	&	0.001	&	0.027	&	0.028	&	0.001	&	0.021	&	0.017	&	0.001	&	0.043	&	0.072	\\
triplet photon sum\tablenotemark{d}	&	0.409	&	-0.042	&	1.211	&	0.426	&	-0.002	&	0.003	&	0.427	&	0.000	&	0.000	\\
average $\chi^2$	&		&		&	63.808	&		&		&	38.399	&		&		&	4.142	\\
\enddata
\tablenotetext{a}{Reddening corrections may be significantly overestimated.}
\tablenotetext{b}{Blended with H$8$.}
\tablenotetext{c}{Reddening corrections may be significantly underestimated.}
\tablenotetext{d}{See text for discussion.}
\label{table:casebcompare}
\end{deluxetable}

\clearpage

\begin{deluxetable}{lllrrrrr}
\tabletypesize{\scriptsize}
\tablecaption{Theoretical and observed relative line intensity ratios of lines with the same upper level.}
\tablewidth{0pt}
\tablehead
{
  \colhead{$n \,{}^{2S+1}\! L$} & 
  \colhead{$n' \,{}^{2S+1}\! L'$} & 
  \colhead{$n'' \,{}^{2S+1}\! L''$} & 
  \multicolumn{2}{c}{Wavelengths (Air)} &
  \multicolumn{2}{c}{$I'/I''$} &
  \colhead{Difference} \\
  \colhead{upper} & 
  \colhead{lower} & 
  \colhead{lower} &
  \colhead{$\lambda'($\mbox{\AA}$)$} & 
  \colhead{$\lambda''($\mbox{\AA}$)$} &
  \colhead{Theor.} & 
  \colhead{Obs.} &
  \colhead{in Std Dev}
}
\startdata
$6d\,{}^{3}\!D$	&	$2p\,{}^{3}\!P$	&	$3p\,{}^{3}\!P$	&	3820	&	10311	&	8.71	&	9.3$\pm$1.5	&	0.4	\\
$7d\,{}^{3}\!D$	&	$2p\,{}^{3}\!P$	&	$3p\,{}^{3}\!P$	&	3705	&	9517	&	8.16	&	23.9$\pm$3.8	&	4.2	\\
$7d\,{}^{1}\!D$	&	$2p\,{}^{1}\!P$	&	$3p\,{}^{1}\!P$	&	4009	&	10138	&	6.68	&	6.3$\pm$1.1	&	-0.3	\\
$7p\,{}^{1}\!P$	&	$2s\,{}^{1}\!S$	&	$3s\,{}^{1}\!S$	&	3355	&	8915	&	10.10	&	11.1$\pm$2.2	&	0.4	\\
$8d\,{}^{3}\!D$	&	$2p\,{}^{3}\!P$	&	$3p\,{}^{3}\!P$	&	3634	&	9063	&	7.86	&	9.5$\pm$1.5	&	1.1	\\
$8p\,{}^{1}\!P$	&	$2s\,{}^{1}\!S$	&	$3s\,{}^{1}\!S$	&	3297	&	8518	&	9.69	&	13.5$\pm$4.8	&	0.8	\\
$9d\,{}^{3}\!D$	&	$2p\,{}^{3}\!P$	&	$3p\,{}^{3}\!P$	&	3587	&	8777	&	7.67	&	4.3$\pm$0.7	&	-5.1	\\
$11p\,{}^{3}\!P$	&	$3s\,{}^{3}\!S$	&	$3d\,{}^{3}\!D$	&	7062	&	8854	&	4.63	&	2.4$\pm$0.5	&	-4.6	\\
$11d\,{}^{1}\!D$	&	$2p\,{}^{1}\!P$	&	$3p\,{}^{1}\!P$	&	3806	&	8931	&	6.15	&	11.0$\pm$3.4	&	1.4	\\
$12p\,{}^{3}\!P$	&	$3s\,{}^{3}\!S$	&	$3d\,{}^{3}\!D$	&	6990	&	8740	&	4.71	&	4.3$\pm$1.3	&	-0.3	\\
$12d\,{}^{3}\!D$	&	$2p\,{}^{3}\!P$	&	$3p\,{}^{3}\!P$	&	3513	&	8342	&	7.39	&	6.0$\pm$1.3	&	-1.1	\\
$12d\,{}^{1}\!D$	&	$2p\,{}^{1}\!P$	&	$3p\,{}^{1}\!P$	&	3785	&	8817	&	6.10	&	7.2$\pm$3.3	&	0.3	\\
$14d\,{}^{3}\!D$	&	$2p\,{}^{3}\!P$	&	$3p\,{}^{3}\!P$	&	3488	&	8204	&	7.30	&	9.7$\pm$2.9	&	0.8	\\
$15d\,{}^{3}\!D$	&	$2p\,{}^{3}\!P$	&	$3p\,{}^{3}\!P$	&	3479	&	8156	&	7.27	&	7.8$\pm$2.4	&	0.2	\\
$16d\,{}^{3}\!D$	&	$2p\,{}^{3}\!P$	&	$3p\,{}^{3}\!P$	&	3472	&	8116	&	7.24	&	10.5$\pm$3.8	&	0.8	\\
$17d\,{}^{3}\!D$	&	$2p\,{}^{3}\!P$	&	$3p\,{}^{3}\!P$	&	3466	&	8084	&	7.22	&	18.0$\pm$10.2	&	1.1
\enddata
\label{table:initiallevel}
\end{deluxetable}

\clearpage

\begin{deluxetable}{lrrr}
\tabletypesize{\scriptsize}
\tablecaption{Average and standard deviation of the helium abundances determined from Table~4 of IT04.}
\tablewidth{0pt}
\tablehead
{
  \colhead{$\lambda$}       & 
  \colhead{$<y^{+}(\lambda)>$}			&
  \colhead{$\sigma$}				\\
}
\startdata
4471 & 0.0778 & 0.0034 \\
5876 & 0.0835 & 0.0033 \\
6678 & 0.0818 & 0.0027 \\
\enddata
\label{table:IT04abundances}
\end{deluxetable}

\clearpage

\begin{deluxetable}{cllcccc}
\tabletypesize{\scriptsize}
\tablecaption{Parameters for fits to He~I emissivities, $4\pi j_{\lambda}/n_{e}n_{\mathrm{He}^+}$.  See text for formula.}
\tablewidth{0pt}
\tablehead
{
  \colhead{Wavelength (Air)}                               & 
  \colhead{$n \,{}^{2S+1}\! L$}                            & 
  \colhead{$n \,{}^{2S+1}\! L$}                            &
  \colhead{}                                              &
  \colhead{}                                              &
  \colhead{}                                              &
  \colhead{}                                              \\
  \colhead{$\mbox{\AA}$}                                   &
  \colhead{upper}                                          &
  \colhead{lower}                                          &
  \colhead{$a$}                                              &
  \colhead{$b$}                                              &
  \colhead{$c$}                                              &
  \colhead{$d$}                                              \\
}
\startdata
2945	&	$5p\,{}^{3}\!P$	&	$2s\,{}^{3}\!S$	&	-1.1849E+05	&	-4.2559E+02	&	1.2913E+04	&	3.5306E+05	\\
3188	&	$4p\,{}^{3}\!P$	&	$2s\,{}^{3}\!S$	&	-2.3591E+05	&	-8.6438E+02	&	2.5969E+04	&	6.9653E+05	\\
3614	&	$5p\,{}^{1}\!P$	&	$2s\,{}^{1}\!S$	&	-3.0442E+04	&	-1.2890E+02	&	3.5599E+03	&	8.5434E+04	\\
3889	&	$3p\,{}^{3}\!P$	&	$2s\,{}^{3}\!S$	&	-6.0645E+05	&	-2.5058E+03	&	7.0181E+04	&	1.7184E+06	\\
3965	&	$4p\,{}^{1}\!P$	&	$2s\,{}^{1}\!S$	&	-6.2095E+04	&	-2.6386E+02	&	7.2700E+03	&	1.7424E+05	\\
4026	&	$5d\,{}^{3}\!D$	&	$2p\,{}^{3}\!P$	&	-4.8009E+04	&	-3.4233E+02	&	7.5849E+03	&	9.2951E+04	\\
4121	&	$5s\,{}^{3}\!S$	&	$2p\,{}^{3}\!P$	&	1.0266E+04	&	6.5524E+01	&	-1.3829E+03	&	-2.6154E+04	\\
4388	&	$5d\,{}^{1}\!D$	&	$2p\,{}^{1}\!P$	&	-1.1349E+04	&	-8.7804E+01	&	1.8895E+03	&	1.9916E+04	\\
4438	&	$5s\,{}^{1}\!S$	&	$2p\,{}^{1}\!P$	&	3.7527E+03	&	2.2646E+01	&	-4.8794E+02	&	-9.9429E+03	\\
4472	&	$4d\,{}^{3}\!D$	&	$2p\,{}^{3}\!P$	&	-3.5209E+04	&	-4.5168E+02	&	8.5367E+03	&	9.1635E+03	\\
4713	&	$4s\,{}^{3}\!S$	&	$2p\,{}^{3}\!P$	&	2.4264E+04	&	1.5802E+02	&	-3.2993E+03	&	-6.1132E+04	\\
4922	&	$4d\,{}^{1}\!D$	&	$2p\,{}^{1}\!P$	&	-6.1979E+03	&	-1.1414E+02	&	2.0047E+03	&	-8.6195E+03	\\
5016	&	$3p\,{}^{1}\!P$	&	$2s\,{}^{1}\!S$	&	-1.3442E+05	&	-5.9029E+02	&	1.6033E+04	&	3.7142E+05	\\
5048	&	$4s\,{}^{1}\!S$	&	$2p\,{}^{1}\!P$	&	7.1813E+03	&	4.5389E+01	&	-9.5007E+02	&	-1.8788E+04	\\
5876	&	$3d\,{}^{3}\!D$	&	$2p\,{}^{3}\!P$	&	2.0620E+05	&	1.7479E+02	&	-1.3548E+04	&	-7.3492E+05	\\
6678	&	$3d\,{}^{1}\!D$	&	$2p\,{}^{1}\!P$	&	6.7315E+04	&	7.5157E+01	&	-4.7101E+03	&	-2.3610E+05	\\
7065	&	$3s\,{}^{3}\!S$	&	$2p\,{}^{3}\!P$	&	5.8675E+04	&	4.1954E+02	&	-8.2053E+03	&	-1.4565E+05	\\
7281	&	$3s\,{}^{1}\!S$	&	$2p\,{}^{1}\!P$	&	1.3544E+04	&	9.9057E+01	&	-1.8831E+03	&	-3.4394E+04	\\
9464	&	$5p\,{}^{3}\!P$	&	$3s\,{}^{3}\!S$	&	-6.5519E+03	&	-2.3532E+01	&	7.1399E+02	&	1.9522E+04	\\
10830	&	$2p\,{}^{3}\!P$	&	$2s\,{}^{3}\!S$	&	-3.9020E+05	&	-1.7846E+03	&	4.9448E+04	&	1.0443E+06	\\
11969	&	$5d\,{}^{3}\!D$	&	$3p\,{}^{3}\!P$	&	-4.8419E+03	&	-3.4526E+01	&	7.6497E+02	&	9.3744E+03	\\
12527	&	$4p\,{}^{3}\!P$	&	$3s\,{}^{3}\!S$	&	-7.5547E+03	&	-2.7681E+01	&	8.3162E+02	&	2.2306E+04	\\
12785	&	$5f\,{}^{3}\!F$	&	$3d\,{}^{3}\!D$	&	3.7439E+04	&	1.0086E+02	&	-3.4590E+03	&	-1.2157E+05	\\
12790	&	$5f\,{}^{1}\!F$	&	$3d\,{}^{1}\!D$	&	1.2480E+04	&	3.3619E+01	&	-1.1530E+03	&	-4.0523E+04	\\
12968	&	$5d\,{}^{1}\!D$	&	$3p\,{}^{1}\!P$	&	-1.4362E+03	&	-1.1111E+01	&	2.3910E+02	&	2.5205E+03	\\
15084	&	$4p\,{}^{1}\!P$	&	$3s\,{}^{1}\!S$	&	-3.3007E+03	&	-1.4026E+01	&	3.8644E+02	&	9.2617E+03	\\
17002	&	$4d\,{}^{3}\!D$	&	$3p\,{}^{3}\!P$	&	-2.4898E+03	&	-3.1940E+01	&	6.0368E+02	&	6.4809E+02	\\
18685	&	$4f\,{}^{3}\!F$	&	$3d\,{}^{3}\!D$	&	8.7712E+04	&	2.8178E+02	&	-8.8285E+03	&	-2.5836E+05	\\
18697	&	$4f\,{}^{1}\!F$	&	$3d\,{}^{1}\!D$	&	2.9217E+04	&	9.3839E+01	&	-2.9404E+03	&	-8.6063E+04	\\
19089	&	$4d\,{}^{1}\!D$	&	$3p\,{}^{1}\!P$	&	-5.7257E+02	&	-1.0544E+01	&	1.8518E+02	&	-7.9601E+02	\\
19543	&	$4p\,{}^{3}\!P$	&	$3d\,{}^{3}\!D$	&	-4.4053E+03	&	-1.6141E+01	&	4.8493E+02	&	1.3007E+04	\\
20580	&	$2p\,{}^{1}\!P$	&	$2s\,{}^{1}\!S$	&	-6.5411E+04	&	-3.4797E+02	&	8.7844E+03	&	1.6613E+05	\\
21118	&	$4s\,{}^{3}\!S$	&	$3p\,{}^{3}\!P$	&	3.7041E+03	&	2.4122E+01	&	-5.0366E+02	&	-9.3323E+03	\\
\enddata
\label{table:emissfits}
\end{deluxetable}

\begin{deluxetable}{lcccc}
\tabletypesize{\scriptsize}
\tablecaption{Fit parameters for the relative collisional contributions, $C/R$, to the emissivities of lines with the given upper level.  See text for formula.}
\tablewidth{0pt}
\tablehead
{
  \colhead{$n \,{}^{2S+1}\! L$}                            & 
  \colhead{}                                              &
  \colhead{}                                              &
  \colhead{}                                              &
  \colhead{}                                              \\
  \colhead{upper}                            & 
  \colhead{$i$}                                              &
  \colhead{$a_i$}                                              &
  \colhead{$b_i$}                                              &
  \colhead{$c_i$}                                              \\
}
\startdata
$3s\,{}^{3}\!S$	&	1	&	37.2702	&	-1.2670	&	-3.3640	\\
	&	2	&	2.6982	&	-1.2918	&	-3.6989	\\
	&	3	&	0.9598	&	-1.3903	&	-4.5122	\\
$3s\,{}^{1}\!S$	&	1	&	17.4945	&	-1.4946	&	-3.5982	\\
	&	2	&	0.5015	&	-1.1724	&	-4.5518	\\
$3p\,{}^{3}\!P$	&	1	&	8.9027	&	-1.0970	&	-3.6989	\\
	&	2	&	1.0004	&	-1.0491	&	-4.3800	\\
	&	3	&	0.5387	&	-0.5550	&	-4.5449	\\
	&	4	&	0.3533	&	-1.1615	&	-4.8186	\\
	&	5	&	0.2345	&	-0.6632	&	-4.9006	\\
$3d\,{}^{3}\!D$	&	1	&	6.7937	&	-0.1116	&	-3.7761	\\
	&	2	&	0.1808	&	-0.8306	&	-4.5122	\\
	&	3	&	1.3478	&	-0.4017	&	-4.5459	\\
	&	4	&	0.4792	&	-0.4062	&	-4.9012	\\
	&	5	&	0.2950	&	-0.8224	&	-4.9013	\\
$3d\,{}^{1}\!D$	&	1	&	0.4340	&	-0.7808	&	-3.7766	\\
	&	2	&	0.1942	&	-0.7687	&	-4.5459	\\
	&	3	&	0.1263	&	-1.1108	&	-4.9012	\\
	&	4	&	0.0549	&	1.8948	&	-4.9013	\\
	&	5	&	0.0785	&	2.0453	&	-5.0942	\\
	&	6	&	0.0935	&	2.0461	&	-5.0942	\\
$3p\,{}^{1}\!P$	&	1	&	2.8310	&	-1.0005	&	-3.7917	\\
	&	2	&	0.8019	&	-1.3326	&	-4.4724	\\
	&	3	&	0.4130	&	-1.0703	&	-4.5452	\\
	&	4	&	0.2782	&	-1.3273	&	-4.8641	\\
	&	5	&	0.1912	&	-1.0948	&	-4.9008	\\
$4s\,{}^{3}\!S$	&	1	&	29.1613	&	-1.3278	&	-4.3800	\\
	&	2	&	1.2121	&	-1.4745	&	-4.5122	\\
$4s\,{}^{1}\!S$	&	1	&	17.4301	&	-1.5851	&	-4.4724	\\
	&	2	&	0.3277	&	-1.3035	&	-4.9043	\\
$4p\,{}^{3}\!P$	&	1	&	8.5965	&	-1.2591	&	-4.5122	\\
	&	2	&	0.6886	&	-1.2274	&	-4.8186	\\
	&	3	&	0.2282	&	-0.7294	&	-4.9006	\\
$4d\,{}^{3}\!D$	&	1	&	4.4397	&	-0.2954	&	-4.5449	\\
	&	2	&	0.1341	&	-0.7331	&	-4.8839	\\
	&	3	&	0.7546	&	-0.5041	&	-4.9012	\\
$4d\,{}^{1}\!D$	&	1	&	2.5507	&	-0.8404	&	-4.5452	\\
	&	2	&	0.3106	&	-0.8657	&	-4.9012	\\
	&	3	&	0.0739	&	2.0732	&	-5.0942	\\
	&	4	&	0.0399	&	2.0304	&	-5.0942	\\
$4f\,{}^{3}\!F$	&	1	&	3.1027	&	-0.2421	&	-4.5459	\\
	&	2	&	0.6790	&	-0.6626	&	-4.9013	\\
	&	3	&	0.0348	&	2.3128	&	-5.0942	\\
	&	4	&	0.0629	&	2.3128	&	-5.0942	\\
$4f\,{}^{1}\!F$	&	1	&	0.9680	&	-0.6405	&	-4.5459	\\
	&	2	&	0.2817	&	-0.9699	&	-4.9013	\\
	&	3	&	0.1045	&	2.3128	&	-5.0942	\\
	&	4	&	0.1885	&	2.3128	&	-5.0942	\\
$4p\,{}^{1}\!P$	&	1	&	2.8275	&	-0.9962	&	-4.5518	\\
	&	2	&	0.5209	&	-1.3550	&	-4.8641	\\
	&	3	&	0.2236	&	-1.1224	&	-4.9008	\\
	&	4	&	0.0641	&	1.8784	&	-5.0730	\\
	&	5	&	0.0385	&	1.8788	&	-5.0940	\\
$5s\,{}^{3}\!S$	&	1	&	33.8477	&	-1.4584	&	-4.8186	\\
	&	2	&	0.4717	&	-1.2851	&	-4.8839	\\
$5s\,{}^{1}\!S$	&	1	&	17.8299	&	-1.5992	&	-4.8641	\\
$5p\,{}^{3}\!P$	&	1	&	5.7783	&	-1.0686	&	-4.8839	\\
$5d\,{}^{3}\!D$	&	1	&	4.0162	&	-0.4399	&	-4.9006	\\
$5d\,{}^{1}\!D$	&	1	&	2.7165	&	-0.9041	&	-4.9008	\\
	&	2	&	0.0841	&	2.0783	&	-5.0942	\\
$5f\,{}^{3}\!F$	&	1	&	3.8606	&	-0.3120	&	-4.9012	\\
	&	2	&	0.0621	&	2.2606	&	-5.0942	\\
$5f\,{}^{1}\!F$	&	1	&	1.5158	&	-0.6943	&	-4.9012	\\
	&	2	&	0.1860	&	2.2606	&	-5.0942	\\
$5p\,{}^{1}\!P$	&	1	&	3.4972	&	-1.0841	&	-4.9043	\\
	&	2	&	0.1006	&	1.8740	&	-5.0730	\\
	&	3	&	0.0413	&	1.8744	&	-5.0940	\\
\enddata
\label{table:collcontrib}
\end{deluxetable}

\end{document}